\documentclass[prl,twocolumn,showpacs]{revtex4}
\usepackage{epsfig}
\usepackage{amsbsy}
\usepackage{amsmath}
\usepackage{latexsym}

\begin{document}

\title{Entanglement versus energy in quantum spin models}
\author{XiaoGuang Wang}
\affiliation{Zhejiang Institute of Modern Physics, Department of
Physics, Zhejiang University, HangZhou 310027, China}

\date{\today}
\begin{abstract}
We study entanglement properties of all eigenstates of the
Heisenberg $XXX$ model, and find that the entanglement and
mixedness for a pair of nearest-neighbor qubits are completely
determined by the corresponding eigenenergies. Specifically, the
negativity of the eigenenergy implies pairwise entanglement. From
the relation between entanglement and eigenenergy, we obtain
finite-size behaviors of the entanglement. We also study
entanglement and mixedness versus energy in the quantum Heisenberg
$XY$ model.
\end{abstract}
\maketitle

Quantum entanglement lies at the heart of quantum mechanics, and
can be exploited to accomplish some physical tasks such as quantum
teleportation~\cite{Tele}. In this sense, it can be regarded as a
{\em resource}, just like energy. As pointed out by Osborne and
Nielsen~\cite{Osborne}, the similarity between entanglement and
energy turns out to be more than superficial. It is interesting to
explore the relationship between these two resources, entanglement
and energy.

Recently, the study of entanglement properties in many-body
systems have received much
attention~\cite{M_NielPOST http://arxiv.org/upload_submit HTTP/ee,M_Jin,M_Kamta,M_OConnor,M_Meyer,M_Brennen,M_Wong,M_Osterloh,M_Sun,QPT_JVidal,QPT_Lambert,QPT_GVidal,QPT_Gu}.
Specifically, for the ground state (zero temperature) of a ring of
$N$ qubits interacting via the antiferromagnetic isotropic
Heisenberg Hamiltonian,  a direct relation is established between
the concurrence $C$~\cite{Conc} quantifying the two-qubit
entanglement and the ground-state energy per site
$\epsilon_0$~\cite{M_OConnor,Victory,Victory2}:
\begin{equation}
C_0(N)=\max[0,-{\cal E}_0(N)/N]=\max[0,-\epsilon_0(N)], \label{ce}
\end{equation}
where $C$ refers to the concurrence for two nearest-neighbor
qubits, and ${\cal E}_0$ is the ground-state energy. For a pair of
qubits, the entanglement of formation can be obtained from the
concurrence $C$
\begin{eqnarray}
E_{\text of}=h\left( \frac{1+\sqrt{1-{C}^2}}2\right) ,
\label{entanglement}
\end{eqnarray}
where $h(x)=-x\log _2x-(1-x)\log _2(1-x).$ The concurrence itself
is a good measure of two-qubit entanglement, and we adopt it as
our measure of pairwise entanglement.

For the case of finite temperature, the concurrence for two
nearest qubits in the thermal state of the ring was found to be
related to the thermodynamical function, the internal energy $U$
via~\cite{Victory,Victory2}
\begin{equation}
C(N)=\max(0,-U/N), \label{ccee}
\end{equation}
which connects the microscopic quantity, the concurrence, and the
macroscopic quantity, the internal energy. Equations (\ref{ce})
and (\ref{ccee}) show that the two-qubit entanglement and energies
are closely related in the isotropic Heisenberg model.

Having known the relations between entanglement and energy for
ground states and thermal states of the $XXX$ model, further
questions arise that what are the entanglement properties of
excited states, and what are the relationships between
entanglement and excited-state energies. We will address the
question in this paper. To study entanglement of excited states is
not only interesting itself, but also help to understand the
entanglement at finite temperatures. We also consider the
entanglement versus energy in the quantum $XY$ model, and show
that the entanglement exhibits a symmetry, distinct from that in
the $XXX$ model.

There exists another concept, the mixedness of a state, related to
entanglement, is also central in quantum information
theory~\cite{Jaeger}. For instance, Bose and Vedral have shown
that entangled states become useless for quantum teleportation on
exceeding a certain degree of mixedness~\cite{Bose}. For a pure
bipartite state, the mixedness of one subsystem is equal to that
of another one, and can be used to quantify bipartite
entanglement. In this case, entanglement and mixedness are
equivalent. We will study both entanglement and mixedness
properties.

{\em Entanglement and eigenenergy.} We consider a physical model
of a ring of $N$ qubits interacting via the isotropic Heisenberg
$XXX$ Hamiltonian
\begin{equation}
H=J\sum_{i=1}^{N-1}S_{i,i+1}+JS_{N,1},  \label{h}
\end{equation}
where $S_{j,j+1}=${}$\frac 12\left( 1+\vec{\sigma}_i\cdot \vec{\sigma}%
_{i+1}\right) $ is the swap operator between qubit $i$ and $j$, $\vec{\sigma}%
_i=(\sigma _{ix},\sigma _{iy},\sigma _{iz})$ is the vector of
Pauli matrices, and $J$ is the exchange constant. Positive and
negative $J$ correspond to the antiferromagnetic and ferromagnetic
case, respectively. Note that we have assumed the periodic
boundary condition.

When an energy level of our system is non-degenerate, the
corresponding eigenstate is pure. When a $k$-th energy level is
degenerate, we assume that the corresponding state is an equal
mixture of all eigenstates with energy ${\cal E}_k$. Thus, the
state correspoding to the $k$-th level with degeneracy becomes a
mixed other than pure, keeping all symmetries of the Hamiltonian.
A degenerate ground state is called thermal ground state in the
sense that it can be obtained from the thermal state $\exp[-H/(k_B
T)]/Z$ by taking the zero-temperature limit~\cite{Osborne}. Here,
$k_B$ is the Boltzman constant, $T$ the temperature, and $Z$ the
partition function. The $k$-th eigenstate $\rho_k$ can be
considered as the thermal ground state of the nonlinear
Hamiltonian $H^\prime$ given by
\begin{equation}
H^\prime=(H-{\cal E}_k)^2.
\end{equation}
Note that Hamiltonian $H^\prime$ inherits all symmetries of Hamiltonian $H$.

Due to the translational invariance of the system, all
nearest-neighbour entanglements are identical. Thus, from now on,
we consider the entanglement between qubit 1 and 2, representing
nearest-neighbour pairwise entanglement. Another important SU(2)
symmetry of the Hamiltonian guarantees that the reduced density
matrix $\rho_{k}^{(12)}=\text{Tr}_{3,4,\cdots,N}(\rho_k)$ for
$k$-th eigenstate is given by
\begin{equation}
\rho^{12}_k=\left(
\begin{array}{llll}
u_{k+} & 0 & 0 & 0 \\
0& u_{k-} & z_k &0  \\
0& z_k & u_{k-} & 0 \\
0& 0 & 0 & u_{k+}
\end{array}
\right)  \label{rho12}
\end{equation}
in the standard basis $\{|00\rangle ,|01\rangle ,|10\rangle ,|11\rangle \}$, with
\begin{equation}
u_{k\pm}=(1\pm G^{zz}_k)/4,\; z_k=G_k^{zz}/2. \label{rho123}
\end{equation}
Here,
$G_k^{\alpha\alpha}=\text{Tr}(\sigma_{1\alpha}\sigma_{2\alpha}\rho_{k}),~\alpha\in\{x,y,z\}$
are correlation functions which are equal due to the SU(2)
symmetry.

{}From Hamiltonian $H$, a simple and useful relation between
eigenenergy per site $\epsilon_k$ and the correlation function is
obtained as
\begin{equation}
\epsilon_k={\cal E}_k/N=\frac {1+3G^{zz}_k}2. \label{rrr}
\end{equation}
Therefore, from Eqs.~(\ref{rho12}-\ref{rrr}), the reduced density
matrix is completely determined by the eigenenergy $\epsilon_k$.
As all information about pairwise entanglement and mixedness of
$\rho^{12}_k$ are contained in $\rho_k^{12}$, the eigenenergy
completely determines the entanglement and mixedness. Next, we
give a quantitative relation between the concurrence and the
eigenenergy.

From the two-qubit reduced density matrix (\ref{rho12}), the
concurrence for two nearest qubits is obtained as~\cite{M_OConnor}
\begin{align}
C_k(N)=&2\max(0,|z_k|-u_{k+})\nonumber\\
=&\max(0,|G_{k}^{zz}|-G_{k}^{zz}/2-1/2)\nonumber\\
=&\max(0, -3G_{k}^{zz}/2-1/2)\nonumber\\
=&\max(0,-\epsilon_k). \label{ce2}
\end{align}
The third equality follows from the inequality $|G^{zz}_k|\le 1$,
which is a special case of a more general result that $|\langle
A\rangle|\le 1$ for any Hermitian operator $A$ satisfying $A^2=1$.
The last equality is obtained by using Eq.~(\ref{rrr}). Thus, we
get a simple relation between the pairwise entanglement of two
nearest qubits and the corresponding eigenenergy. A necessary
condition for non-zero entanglement is that eigenenergy is
negative, i.e., the negativity of eigenenergy implies pairwise
entanglement. We also see that the ground state exhibits largest
entanglement.

Now we apply the above result to investigate finite-size
behaviours of the entanglement. Conformal invariance theory
predicts the finite size behaviour of the ground state and the
first excited state~\cite{Conformal}
\begin{equation}
\epsilon_1(N)-\epsilon_0(N)\approx \pi^2 b/N^2 \label{conf}
\end{equation}
with $b$ being identified with the central charge. From the above
equation and Eq.~(\ref{ce2}), we immediately have
\begin{equation}
C_1(N)-C_0(N)\approx -\pi^2 b/N^2, \label{conff}
\end{equation}
which is the finite size behaviour of the entanglement.

In Table I, we give the concurrences of the ground states, first
and second excited states for the number of qubits from 3 to 16.
The ground state becomes entangled when $N\ge
4$~\cite{Victory,Victory2}. However, the first (second) excited
state becomes entangled when $N\ge 6$ ($N\ge 7$). The concurrence
of the ground state decreases (increases) as even $N$ (odd $N$)
increases. However, the concurrence of the first excited states
increases as even $N$ or odd $N$ increases. This fact also holds
for the second excited states. We see that the entanglement
properties of the excited states are distinct from those of the
ground states.

\begin{table}[tbp]
\caption{The concurrences of the ground state, first and second
excited states for the number of qubits from 3 to 16.}
\begin{tabular}{ccccccccccc}
$N$       & 3   & 4    & 5      & 6      & 7      &8     &9
\\ \hline
$C_{0}$ &0.0  & 0.5  & 0.2472 & 0.4343 & 0.3158 &0.4128&0.3438
\\ \hline
$C_{1}$ &0.0  & 0.0  & 0.0    & 0.2060 & 0.0161 &0.2821&0.1525
\\ \hline
$C_{2}$ &-    & 0.0 & 0.0     & 0.0    & 0.0055 &0.1749&0.1342
\\ \hline
$N$     & 10    & 11      & 12    & 13    & 14     & 15    &16
\\ \hline
$C_{0}$ & 0.4031& 0.3580  & 0.3979& 0.3661& 0.3948 & 0.3712&0.3928
\\ \hline
$C_{1}$ & 0.3184& 0.2258  & 0.3386& 0.2695& 0.3509 & 0.2976&0.3590
\\ \hline
$C_{2}$ & 0.2541& 0.2100  & 0.2962& 0.2570& 0.3212 & 0.2877&0.3371
\end{tabular}
\end{table}

To clearly display behaviours of the pairwise entanglement
corresponding to different energy levels, in Fig.~1, we plot the
concurrences versus energy level index $k$ for different $N$. It
is evident that for a fixed $N$ the concurrence decreases
monotonically with the increase of $k$, and when $k$ is equal or
larger than a threshold value $k_\text{th}$, the entanglement
vanishes. The threshold values $k_\text{th}$ is $N$-dependent, and
the larger the number of qubits, the larger the threshold value.
More and more excited states become entangled when the number of
qubits increases.

\begin{figure}
\includegraphics[width=0.45\textwidth]{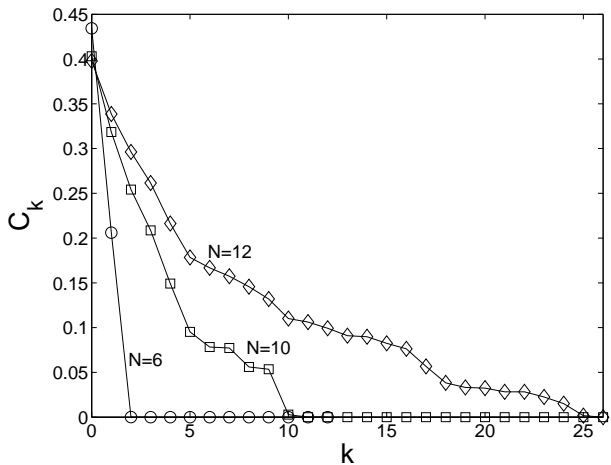}
\caption{The concurrence versus energy level index $k$ for
different number of qubits.}
\end{figure}

\begin{figure}
\includegraphics[width=0.45\textwidth]{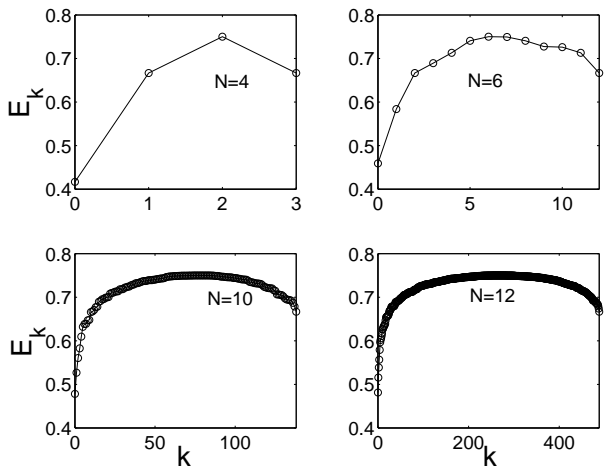}
\caption{The linear entropy versus $k$ for different $N$.}
\end{figure}

Next, we study the mixedness properties of all eigenstates. The
mixedness of a state $\varrho$ can be quantified by the linear
entropy given by
\begin{equation}
E=1-\text{Tr}(\varrho^2).\label{mixed}
\end{equation}
Then, from Eqs.~(\ref{rho12}), (\ref{rho123}), (\ref{rrr}), and
(\ref{mixed}), the linear entropy of the two-qubit state
$\rho_k^{12}$ is obtained as
\begin{equation}
E_k(N)=1-\frac{1}{3}\left[\epsilon_k^2(N)-\epsilon_k(N)+1\right],\label{ce3}
\end{equation}
which is determined solely by the eigenenergy $\epsilon_k$ per
site. Obviously, the linear entropy takes its maximum 3/4 when
$\epsilon_k=1/2$.

In Fig.~2, we plot the linear entropy versus $k$ for different
number of qubits. As $k$ increases, the linear entropy
monotonically increases and reaches a maximum, then monotonically
decreases. In contrast to the concurrence, the linear entropy
takes its maximum in the middle of energy levels, other than at
the ground-state level. For instance, when $N=6$, there are 13
levels and the linear entropy takes its maximum when $k=6$. There
are some eigenstates which are non-degenerate, and thus pure. In
this case, the entanglement of between the two qubits and the rest
quantified by the linear entropy is equivalent to the mixedness of
the two-qubit state. For instance, when $N=6$, the ground state,
the second excited state, and the 10-th excited state are
non-degenerate.

In the isotropic Heisenberg model described above, both
entanglement and mixedness of an eigenstate are completely
determined by the corresponding eigenenergy. This result is due to
the many symmetries in the model. For other models, the relation
between entanglement and energy may become more complicated. We
now consider another well-established model, i.e., the Heisenberg
$XY$ model. The $XY$ Hamiltonian is written as
\begin{equation}
H_{\text{XY}}=\frac{J}2\sum_{i=1}^N
(\sigma_{ix}\otimes\sigma_{i+1x}+\sigma_{iy}\otimes\sigma_{i+1y}),
\end{equation}
where the periodic boundary condition is assumed again.

\begin{figure}
\includegraphics[width=0.45\textwidth]{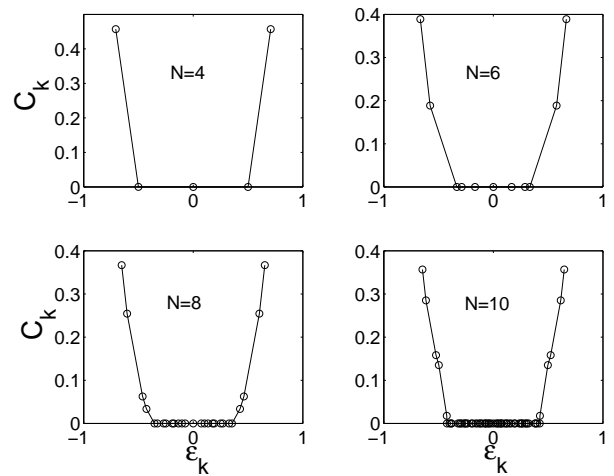}
\caption{The concurrence versus eigenenergy per site $\epsilon_k$
for different number of qubits in the $XY$ model.}
\end{figure}

An interesting feature of this system is that for even $N$ if
${\cal E}$ is an eigenenergy, $-{\cal E}$ is an eigenenergy too.
Thus, the eigenspectrum exhibits a symmetry, which is different
from that in the $XXX$ model. The reason is that for even $N$
Hamiltonian $H_{\text{XY}}$ anticommutes with the following
operator
\begin{equation}
\Lambda_z=\sigma_{1z}\otimes\sigma_{3z}\otimes\cdots\otimes\sigma_{N-1z}.
\end{equation}
It is natural to ask if the entanglement and mixedness exhibit a
similar symmetry. The answer is yes as we will see below.

By exact diagonalization method, we compute the concurrence and
mixedness numerically in the $XY$ model. In Fig.~3, we plot the
concurrences for two nearest qubits versus eigenenergy per site
$\epsilon_k$. It is evident that the concurrence shows a symmetry
with respect to the point of $\epsilon_k=0$, which arises due to
the fact that the operator $\Lambda_z$ is a local unitary
operator, and the entanglement properties of the eigenstate with
energy ${\cal E}$ are the same as those of the eigenstate with
energy $-{\cal E}$. If we restrict ourselves to the half part of
the eigenspectrum ($\epsilon_k\le 0$), the behaviour of the
concurrence is similar to that in the $XXX$ model, i.e., the
concurrence monotonically decreases as $k$ increases and there
also exist threshold values $k_{\text{th}}$.

\begin{figure}
\includegraphics[width=0.45\textwidth]{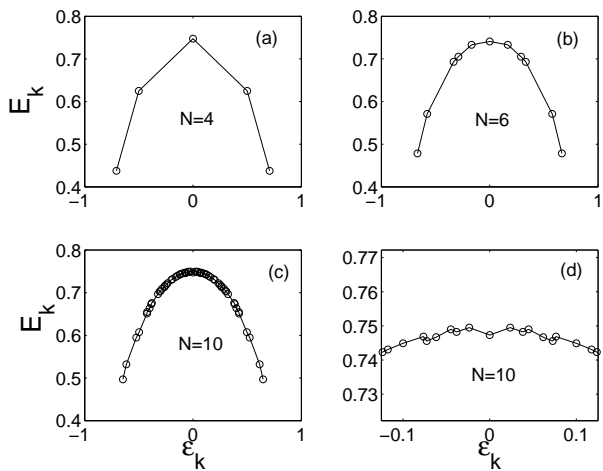}
\caption{The linear entropy versus eigenenergy per site
$\epsilon_k$ for different $N$ in the $XY$ model.}
\end{figure}

The linear entropy quantifying the mixedness of the two-qubit
states is plotted in Fig.~4. It also exhibits a symmetry with
respect to $\epsilon_k=0$. And it seems that the linear entropy
takes its maximum at $\epsilon_k=0$ for any $N$ from Fig.4 (a-c).
However, the partly enlarged version (Fig.~4 (d)) of Fig.~4 (c)
clearly shows that the maximum is not exactly at $\epsilon_k=0$
for $N=10$. So, we see that the mixedness reaches its maximum at
or near the point of $\epsilon_k=0$.

In conclusion, we have studied the entanglement and mixedness of
all eigenstates in two well-established quantum spin models, i.e.,
the $XXX$ and $XY$ models. For the $XXX$ model, the entanglement
and mixedness of a pair of nearest-neighbor qubits in a eigenstate
are completely determined by the corresponding eigenenergy. The
entanglement decreases as we go from ground state to excited
states, i.e. the more excited the system, the less the
entanglement. The negativity of eigenenergy implies pairwise
entanglement, i.e., the entanglement vanishes when the eigenenergy
is large than zero. In contract to the entanglement property, the
mixedness exhibits a maximum in the middle of energyspectrum,
other than on the border. From the relation between entanglement
and eigenenergy, we obtain the finite-size behaviors of the
entanglement.

The properties of entanglement and mixedness in the $XY$ model are
distinct from those in the $XXX$ model. For even number of qubits,
they both exhibits a symmetry with respect to the zero
eigenenergy. However, for odd number of qubits, the symmetry
breaks. It will be interesting to consider entanglement of all
eigenstates in other physical systems such as quantum chaos
system. The study of eigenvalue statistics and eigenvectors
statistics naturally motivate us to study entanglement statistics,
which is under consideration.

\end{document}